\documentclass[manuscript]{aastex}

\newcommand{\ms}{\mbox{m\,s$^{-1}$}}

\newcommand{\Msun}{\mbox{M$_{\odot}$}}
\newcommand{\Rsun}{\mbox{R$_{\odot}$}}

\newcommand{\Lsun}{\mbox{L$_{\odot}$}}
\newcommand{\Mearth}{\mbox{M$_{\oplus}$}}
\newcommand{\Rearth}{\mbox{R$_{\oplus}$}}

\shorttitle{Three planets around GJ628}
\shortauthors{Wright et al.}

\begin{document}


\title{Three planets orbiting Wolf 1061}


\author{D.J. Wright, R.A. Wittenmyer, C.G. Tinney, J.S. Bentley, and 
Jinglin Zhao}
\affil{Department of Astronomy and Australian Centre for Astrobiology, School of Physics, University of New South Wales,
    NSW 2052, Australia}
\email{duncan.wright@unsw.edu.au}

\begin{abstract}
We use archival HARPS spectra to detect three 
planets orbiting the M3 dwarf Wolf 1061 (GJ 628).  We detect 
a 1.36\,\Mearth\ minimum-mass 
planet with an orbital period $P$ = 4.888\,d (Wolf\,1061b), a 4.25\,\Mearth\ minimum-mass planet with orbital period $P$ = 17.867\,d (Wolf\,1061c), and a likely 5.21\,\Mearth\ minimum-mass planet with orbital period $P$ = 67.274\,d (Wolf\,1061d). All of the planets are of 
sufficiently low mass that they may be rocky in nature.  The 
17.867\,d planet falls within the habitable zone for Wolf 1061 and the 
67.274\,d planet falls just outside the outer boundary of the habitable 
zone.  There are no signs of activity observed in the bisector 
spans, cross-correlation full-width-half-maxima, Calcium H \& K 
indices, NaD indices, or H$\alpha$ indices near the planetary periods.  We use custom methods to 
generate a cross-correlation template tailored to the star. The resulting velocities do not suffer the strong annual variation observed in the HARPS DRS velocities. This differential technique should deliver better 
exploitation of the archival HARPS data for the detection of planets at extremely low amplitudes.
\end{abstract}


\section{Introduction}

The NASA {\em Kepler} mission has demonstrated that a high fraction of 
low-mass stars are planet hosts \citep{Dre2013, Dre2015, Kop2013b}, and the 
evidence is mounting that multi-planet systems are common with 
discoveries like GL\,581 \citep{Udry2007}, GL\,667C \citep{Del2013}, and GL\,876 \citep{Riv2005}.
Multiple long-term programs have
demonstrated the utility of Doppler techniques for exoplanet detection
and characterisation (e.g., the HARPS, CORALIE and ELODIE groups, the
Anglo-Australian Planet Search group, the Keck HiRes and Lick Observatory groups).
Several new instruments are being built to expand the use of
Doppler techniques, with a focus on M dwarf planet hosts
(e.g. CARMENES, \cite{Alo2015}; SUBARU HDS, \cite{Sne2015}; ESPRESSO \cite{Pep2014}; {\em Veloce\footnote{http://newt.phys.unsw.edu.au/$\sim$cgt/Veloce.html}}).  

M dwarfs are good Doppler targets due to their many 
sharp molecular absorption features, and their typically slow rotation speeds.  
Low-mass stars are also low-luminosity stars, which contracts their habitable zones 
to short periods (typically \textless 100\,d).  These close-in planets have a stronger Doppler effect on 
their host star and that Doppler amplitude is further increased by the low mass 
of the host star.  All of these characteristics make M dwarfs excellent targets for rocky, habitable-zone exoplanet searches e.g. \citet{Ber2015}.

The 2017 launch of the NASA \textit{Transiting Exoplanet Survey Satellite} (TESS) 
mission \citep{ricker15} will deliver a
quantum leap in the number of M dwarf planet candidates for Doppler follow-up -- many of them in the habitable zones of their hosts. 

The HARPS spectrograph was purpose-built for planet hunting 
\citep{mayor03}.  It is one of the most precise exoplanet hunting facilities 
with a demonstrated long 
term precision of \textless\,1\,m\,s$^{-1}$
\citep{Pep2003,Dum2015}.  The extensive HARPS 
database contains long-term monitoring data for more than one hundred M dwarfs and 
is an excellent resource for the refinement of planet search techniques.

As noted in \cite{Dum2015}, the HARPS DRS velocities suffer from a 
significant yearly variation due to the barycentric movement of the 
spectra across pixel discontinuities in the HARPS detector.
A quick analysis of the HARPS DRS data shows that this annual signal
is clearly the largest variation present for Wolf 1061.  

We have developed a process to extract Doppler velocities from 
the DRS spectra with improved precision over the standard  
analysis using a cross-correlation template {\em generated from the data
itself}, rather than from {\em ab initio} line lists.  The template generated
is therefore inherently differential. We obtained a typical velocity precision of 
\textless\,1\,m\,s$^{-1}$. Most interestingly, analysis using
this template delivers velocities that do not show any significant yearly 
or half yearly periodicities for Wolf 1061.

The observations and methods used to determine Doppler velocities of Wolf 1061 are outlined in sections 2 and 3, while section 4 discusses the velocity and stellar activity results and detected planetary signals. Section 5 focuses on details of the star's habitable zone, and section 6 provides a brief conclusion of our main results.

\section{Observations}

We have used the 148 publicly available HARPS spectra of Wolf 1061. 
The spectra span 380\,\textendash\,680\,nm at resolution $\sim$110\,000.  The observations were taken over 10.3 years and typically have 900s exposure times and a median signal-to-noise (S/N) of 65/pixel at 6000\,\AA.

In the most recent HARPS M dwarf 
sample publication \citep{Bon2013} Wolf 1061, though 
analysed with a much smaller number of observations than presented here, was not considered exceptional.  No significant periodicities were detected, though a test for variability indicated possible variation 
at 67.3d, which we identify as a likely planet orbit.

\section{Obtaining Doppler velocities}

Details on the calibration and precision from the HARPS DRS 
reduced data are discussed in \cite{Pep2003}.  We use the archival spectra of Wolf 1061 in their 72 order, 
flux-per-camera-pixel form i.e. not in their merged, flattened and 
rebinned form. Our method of extracting velocities is similar to the 
HARPS DRS except for three changes:\newline
-	an additional cleaning of outlying pixels in the spectra,\newline
-	a custom template built for the star,\newline
-	an iterative solution using the measured velocities to improve the custom template.\\

As in the standard HARPS analysis \citep{Pep2002}, we determine a velocity from each echelle-order spectrum in each observation by cross-correlation against a weighted spectral mask.
However, rather than using an {\em ab initio} line list, we construct a custom mask for each target star using a high S/N spectrum, which we obtain from the data for that star. 

First, all spectra are shifted to the barycentric reference frame and then any known Doppler velocity is removed. In the first iteration the Doppler velocity is unknown and so initially the HARPS DRS velocities are used. In later iterations the computed velocity is used instead. Errors in the spectra are estimated as the square-root of the flux at each pixel. All spectra are then flattened by the spectrograph blaze function (defined as part of the DRS reduction) and rebinned onto a standard axis of 0.01\,\AA\ bins. The spectra are then inspected to identify and correct outliers at the 5$\sigma$~level using the ensemble of data at each wavelength.

Next a very high S/N spectrum is obtained from summing all the rebinned spectra. To construct the mask, this summed spectrum is upsampled to 0.001\,\AA\ and inverted. The inversion is achieved by assuming a flat pseudo-continuum at the maximum value in the summed spectrum and subtracting the spectrum from the pseudo-continuum value. This allows precise identification of the depth (now seen as line ``height'' due to the inversion) and position of all of the pseudo-line absorption profiles present in the summed spectrum. A cross-correlation template can now be constructed from the list of positions and depths so identified.

To obtain Doppler velocities the spectra go through a similar process to the above (i.e. shifting and cleaning), though without the subtraction of the Doppler velocity. The spectra are then rebinned to a logarithmic wavelength scale in 300\,m\,s$^{-1}$ bins. The spectral template for each order is built using delta-functions with the positions and depths obtained above and covers the spectrum up to 2\,\AA\ from the ends of each order. There are four wavelength regions considered too contaminated by telluric absorption to be useable for this method: 5850\,\textendash\,6030\,\AA; 6270\,\textendash\,6340\,\AA; 6450\,\textendash\,6610\,\AA; 6855\,\AA\,\textendash\ 6910\,\AA. Finally, the template and spectra are cross-correlated and the cross-correlation profiles for each order are added. The S/N in the summed spectrum is insufficient to construct a reliable template for Wolf 1061 at the blue wavelengths, so the bluest 25 orders are not included in the combined cross-correlation profile. 

Velocities are computed by fitting a Gaussian to the combined cross-correlation profile. The velocities are then fed back into the building process for the cross-correlation template as mentioned above. Iterations continue until the difference between velocities from iterations is less than 20\,cm\,s$^{-1}$.

We compute our Doppler velocity errors in a similar manner to the 
DRS system and so obtain similar error estimates, 
typically \textless\,1m\,s$^{-1}$.  This error is based on the 
slope and flux present in the combined cross-correlation profile and
calculated in the same way as outlined by 
\cite{But1996}. Figure 1 shows all of the resulting Doppler velocities. 

The Doppler velocities obtained from the above method do not suffer the annual variation observed in the DRS velocities. By building the template from the shifted and combined spectra the impact of the `seams' in the CCD \citep{Dum2015} are built into the template. The template we construct is an inherently differential measurement and therefore is less sensitive to the causes of the yearly variation in the DRS velocities, which cross-correlates spectral data with an imperfect wavelength solution against an \textit{ab initio} line list with a ``perfect'' wavelength solution. A following paper will discuss our method in more detail to investigate this in greater depth.

\section{Results}

We begin with a sequential extraction of periodicities using the Lomb-Scargle periodogram \citep{Lom76,Sca82} on the full velocity data set. The significant signals found are used as start points for a genetic algorithm analysis of the velocity data. The top panel of
Figure~\ref{planets} shows the periodogram
of the full data set and the location of the three significant periods detected.  At each stage, we fit a Keplerian
to the data with a starting period corresponding to the strongest periodogram peak.
After the subtraction of these three signals no significant variation is 
observed in the residuals, which have an rms of 1.86\,m\,s$^{-1}$.  We 
determined the false-alarm probability (FAP) of each signal using a 
bootstrap randomization approach \citep{kurster97} coupled with the 
error-weighted generalized Lomb-Scargle periodogram \citep{zk09}.  We 
used 10,000 bootstrap realizations to derive the following FAPs for the 
three signals: 17.87\,d -- $<0.0001$, 67.27\,d -- $<0.0001$, 4.89\,d -- 0.0052.

A variety of phenomena including stellar 
activity and stellar pulsation can induce signals in the RVs \citep[e.g.][]{Que2001, robertson14}. Many early M dwarfs have long rotation periods ($>$20\,d) and also have activity e.g. starspots or magnetic effects, that cause variations in the stellar spectrum that are detected as Doppler velocity changes and hence the rotation period or its harmonic may be misinterpreted as a planet. To identify the detected velocity signals in Wolf 1061 as planets we must rule out the possibility that they are caused by intrinsic variability in the host star or that they are associated with the rotation period of the star. 

By following \cite{Sua15} we can determine the log$_{10}$(R$'_{HK}$) index for Wolf 1061 from the Ca\,II H\&K lines which then provides an estimate of the rotation period. \cite{Sua15} define the so-called S-index using a 0.4\,\AA\ window around 3933.664\,\AA\ and 3968.470\,\AA\ to define the Ca\,II H\&K emission and a 20\,\AA\ window around 3901.070\,\AA\ and 4001.070\,\AA\ for the continuum definition. The computation of log$_{10}$(R$'_{HK}$) from the S-index was then straight forward using $B-V$ = 1.566 for Wolf 1061 from \cite{Lan97}. We obtain log$_{10}$(R$'_{HK}$) = -6.00\,$\pm$\,0.13 indicating Wolf 1061 is extremely inactive. \cite{Sua15} provide an equation for the relationship between the rotation period and the log$_{10}$(R$'_{HK}$) value but it is poorly constrained for M dwarf stars and was compiled from a set of stars that only covered values down to log$_{10}$(R$'_{HK}$) = -5.98. Although the relationship provides a value for the rotation period of 199\,d, due to its uncertainty we take from this calculation only that the measured log$_{10}$(R$'_{HK}$) is indicative of a long rotation period e.g. $>$ 100\,d.

Additionally there were 537 epochs (8.6 yr) of precision (error $<$ 4\,mmag) $V$ band photometry available from the All-Sky Automated Survey\footnote{http://www.astrouw.edu.pl/asas} \citep{asas} photometry. Examination of the periodogram of the photometry revealed no significant periodicities (FAP $<$ 0.1) at any period. This is further evidence in support of a rotational period longer than our detected RV periods for Wolf 1061.

Recent work by \cite{Rajpaul2016} has also shown that small peaks in the window function due to the data sampling can masquarade as significant signals after the subtraction of any detected activity signals. Essentially, the subtraction of signals from a dataset can suppress other peaks in a power spectrum and leave behind spurious signals. Therefore we examine the following activity periodograms in Figure~\ref{planets} directly without subtraction.

In addition to the S-index outlined above, there are several other indices known to vary with stellar activity or 
pulsation: the bisector span, the cross-correlation function 
full-width-at-half-maximum (FWHM), H$\alpha$ index and  Na\,D line index \citep{Vau78,San02,Hat2010,Sua15}.
Any significant variation observed 
in these indices at periods coinciding with detected signals in our 
velocities would suggest that signal is likely due to stellar 
activity, pulsation or rotation, and not an exoplanet.

We have used a typical bisector span measurement i.e. the difference 
between the average of the 10-40\% and 60-90\% depths of the bisectors \citep{Pep2003}. 
The H$\alpha$ index emission region was difficult to define because the H$\alpha$ line is weak and there was very little variation observed in the spectra. We settled on using a 2\,\AA\ window centred at 6562.810\,\AA\ for the H$\alpha$ line and a 4\,\AA\ 
window centred on 6556.000\,\AA\ for the continuum. 
We note that the so called `continuum' as used here is not pure continuum emission (pure continuum is not easily accessible in M Dwarf stars) but rather just a region of stable spectrum that should not vary under the same circumstances that the H$\alpha$ line varies e.g. due to starspots or magnetic effects. In this way we measure the strength of the H$\alpha$ emission and its changes relative to a stable region of spectrum.
For the NaD index we used 0.4\,\AA\ window around the D lines at 5895.924\,\AA\ and 5889.954\,\AA, and a 14\,\AA\ window centred on 5875.000\,\AA\ for the continuum region. 
The FWHM of the cross-correlation profile and its error are a direct result of the Doppler velocity measurement process.

In Figure~\ref{planets} the window function for the periodograms shows no peaks at our velocity signal periods but does have two small peaks at 75\,d and 92\,d and shows that our long period sampling $>$ 300\,d is very poor. 
In Figure~\ref{planets} panels 3 - 7 we show periodograms of the activity indices with the positions of the three signals found in our velocity data overplotted.

There are no indications of significant signals in any of the indices at the detected velocity periods.
In fact only the S-index shows any significant signals at periods $<$ 300\,d.  The S-index shows several significant peaks at periods $>$ 75\,d. We interpret the long period peaks ($>$ 300\,d) in all of the activity indices as likely to be due to the data sampling, since the window function clearly indicates that we are not sensitive to such periods. The several significant S-index peaks at $>$ 75\,d may be associated with the rotation period of Wolf 1061 but it is not possible to identify what that period might be. Although there is no direct link between this S-index variability and our RV signal periods, it is impossible to completely rule out a connection between our longest RV period (67.2\,d) and the S-index signals since rotation related velocity signals could show up as harmonics of the rotation period i.e. $P_{\textrm{rot}}$/2 or $P_{\textrm{rot}}$/3.

We also check for correlations between the velocities and activity indices and obtain correlation coefficients ranging from -0.11 to +0.12. We use a simple two-tailed $t$-test and find no significant correlations between the velocities and any of the indices examined here (significant p-values are $p <$ 0.05).

Based on these tests Wolf 1061 is a very stable star and hence the signals detected in the RVs are best explained as planets.
Taking the RV datasets periodicities as starting points, we explored a broad parameter space using a genetic algorithm to fit three Keplerian 
orbits to the data \citep[e.g.][]{NNSer, songhu}.  Approximately $10^7$ 
three-planet models were explored, and convergence occurred rapidly, 
favouring a system with three low-mass, low-eccentricity planets.  We 
derived final system parameters using \textit{Systemic 2} 
\citep{mes09} and estimated parameter uncertainties using the bootstrap 
routine therein on 10,000 synthetic data set realisations.  The 
parameters given in Table~\ref{planetparams} represent the median of the 
posterior distribution and 68.7\% confidence intervals.  Using the 
host star mass of 0.25\,\,\Msun\ \citep{Mal2015}, we derive planetary 
minimum masses of 4.25$\pm$0.37\,\,\Mearth~\ (planet c), 
5.21$\pm$0.68\,\,\Mearth~\ (planet d), and 1.36$\pm$0.23\,\,\Mearth~\ (planet 
b).  Figure~\ref{phaseplots} shows the velocity time-series phased to the 
period of each individual signal with the other planets removed.

We have detected three signals in the velocity timeseries at $P=17.867$\,d, $P=67.274$\,d and $P=4.888$\,d. The variability observed in the activity S-index is at longer periods than our detected velocity signals and does not show any clear association with them, though we acknowledge a possible association with our longest period velocity signal of $P=67.274$\,d cannot be completely ruled out - for the present we cautiously interpret this signal as an exoplanet.

Our velocities and activity indices and the HARPS DRS velocities are available in Table~\ref{starRVs}.

\section{The host star and the habitable zone}

Wolf 1061 is a bright and close M dwarf with V$=10.1$, d$ = 
4.29\,$pc, and spectral type M3/3.5V \citep{Bon2013,Mal2015}.  \cite{Bon2013} provided
several fundamental parameters for Wolf 1061, though a recent study by \cite{Mal2015} has reported it as being slightly cooler than \cite{Bon2013}.  The stellar parameters taken for the computation of the mass of the planets 
and the habitable zone of the star are those of \cite{Mal2015}: 
effective temperature $T_{\textrm{eff}}$\,=3393\,K, mass 
$M$\,=0.25\,\Msun\ and luminosity 
$L$\,=0.007870\,\Lsun.

We use the habitable zone calculations of \citet{Kop2013a} and 
\citet{Kop2014}\footnote{available online at 
\url{http://depts.washington.edu/naivpl/sites/default/files/index.shtml}}, which 
give both conservative and optimistic habitable zones (depending on the approach accepted).  
We obtain 
conservative habitable zone limits of 0.092--0.18\,AU and 
optimistic limits of 0.073--0.19\,AU.  These boundaries 
place the middle planet well inside the optimistic habitable zone, and 
just outside the inner boundary of the conservative habitable zone.  The 
outer planet in the system lies just outside the outer boundary of the 
optimistic habitable zone for the host star.

\section{Conclusions}

We have found strong Doppler signals in data for Wolf 1061 that 
indicate the presence of three potentially rocky planets: a 1.36\,\Mearth\ minimum-mass 
planet with an orbital period of 4.888\,d (Wolf\,1061b), a 4.25\,\Mearth\ minimum-mass planet with orbital period $P = 17.867$\,d (Wolf\,1061c), and a probable 5.21\,\Mearth\ minimum-mass planet with orbital period $P=67.274$\,d (Wolf\,1061d).  With such short-period planets, we can consider the possibility that one or more may transit.  The 
\textit{a priori} transit probabilities are as follows: planet b -- 
14.0\%, planet c -- 5.9\%, planet d -- 2.6\%.  If the planets are transiting, the planet masses are equal to the minimum-masses above. We can then estimate planetary 
radii from the mass-radius relation of \citet{weiss14}.  This yields the 
following rough radius estimates: 1.44\,\,\Rearth\ (b), 
1.64\,\,\Rearth\ (c) and 2.04\,\,\Rearth\ (d).  Using the stellar radius 
given in \cite{Mal2015} ($R=0.26$\,\Rsun) we estimate transit 
depths of 2.6, 3.3, and 5.2 mmag for planets b, c, and d, respectively.  
Such signals can be detected from ground-based telescopes in excellent 
conditions, and we encourage purpose-built facilities like MEarth and 
MINERVA \citep{deeznuts08, swift15} to pursue the Wolf 1061 transit 
windows when this target becomes observable again in early 2016.

The 17.867\,d planet is of particular interest because it is of 
sufficiently small mass to be rocky and is in the habitable zone of the 
host M dwarf star.  The probable outer planet at 67.274\,d resides just on the 
outer boundary of the habitable zone and is also likely rocky.  These 
planets join the small but growing ranks of potentially habitable rocky 
worlds orbiting nearby M dwarf stars.

\acknowledgments

Based on data obtained from the ESO Science Archive Facility under 
request: Duncan Wright 189972. This research has been supported by ARC Super Science Fellowships FS100100046
and ARC Discovery grant DP130102695. 
We also wish to acknowledge the many helpful comments from the referee in producing this manuscript.

\clearpage

\begin{figure}
\plotone{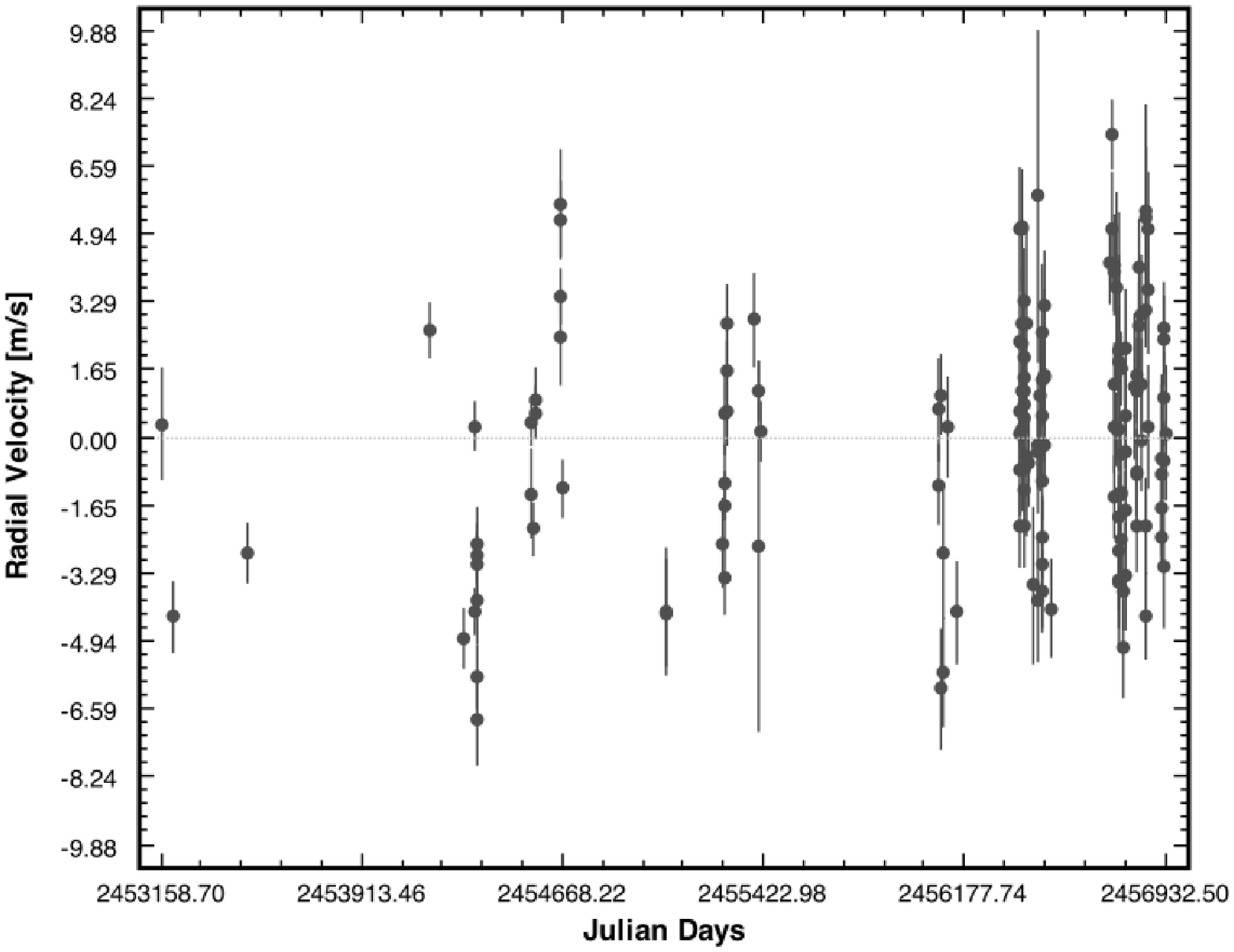}
\caption{The complete set of Doppler velocities for Wolf 1061 that result from the processing
of HARPS data with our differentially generated mask.}
\label{vels}
\end{figure}

\clearpage

\begin{figure}
\epsscale{1.0}
\plotone{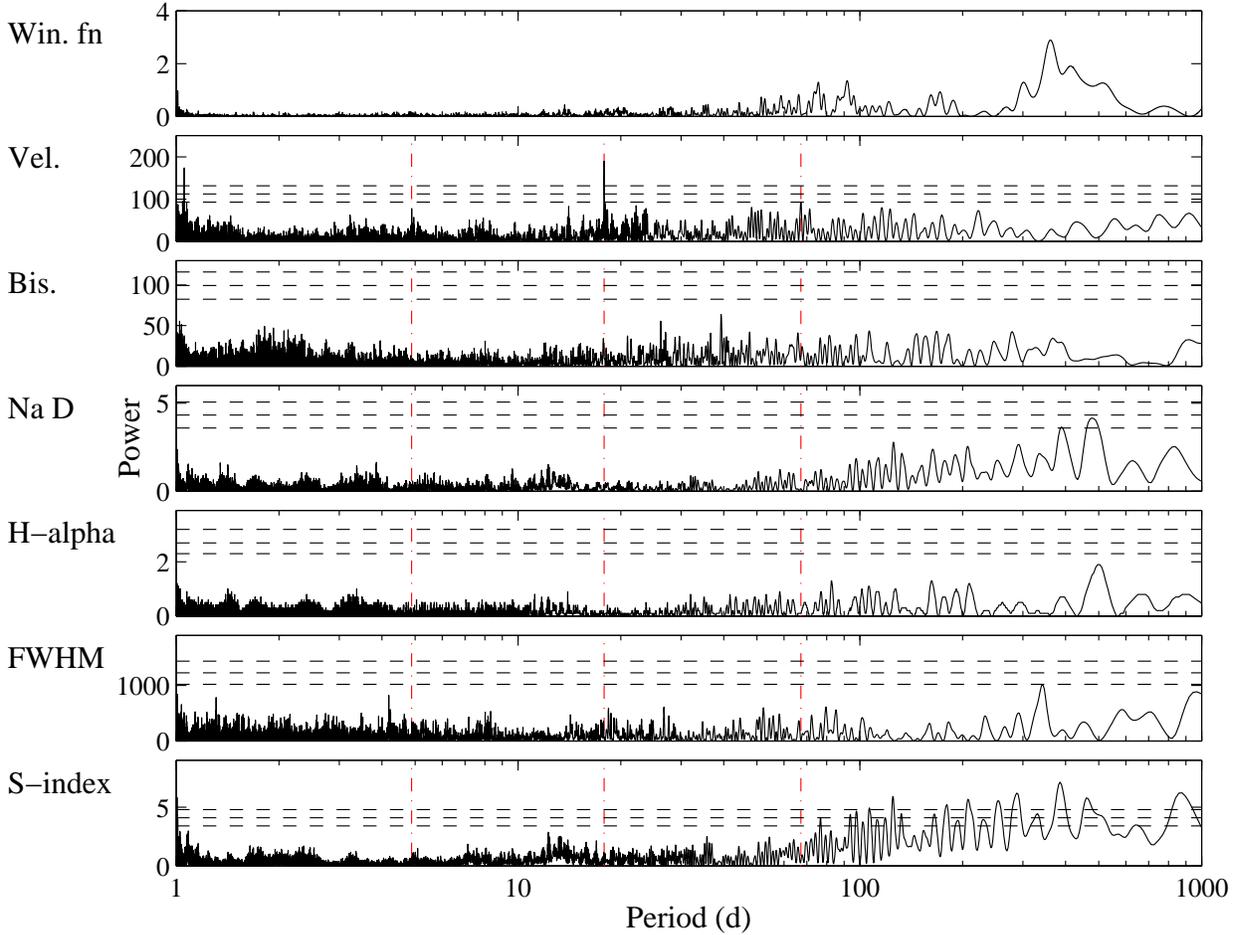}
\caption{The top panel shows the window function for our data set. The second panel is the 
Lomb-Scargle periodogram of the Doppler velocities. The positions of the significant periodicities found in the velocities are marked with dotted vertical lines in panels 2-7. Panels 3-7 show the periodograms of the activity indices. The dashed horizontal lines show the levels corresponding to 10\%, 1\% and 0.1\% false-alarm probabilities using the method of \cite{kurster97}. The H$\alpha$~index and the S-index periodograms have been scaled by 100\,000 and 1\,000 respectively because the index values and variation are so small.
}
\label{planets}
\end{figure}

\clearpage

\begin{figure}
\epsscale{0.85}
\plotone{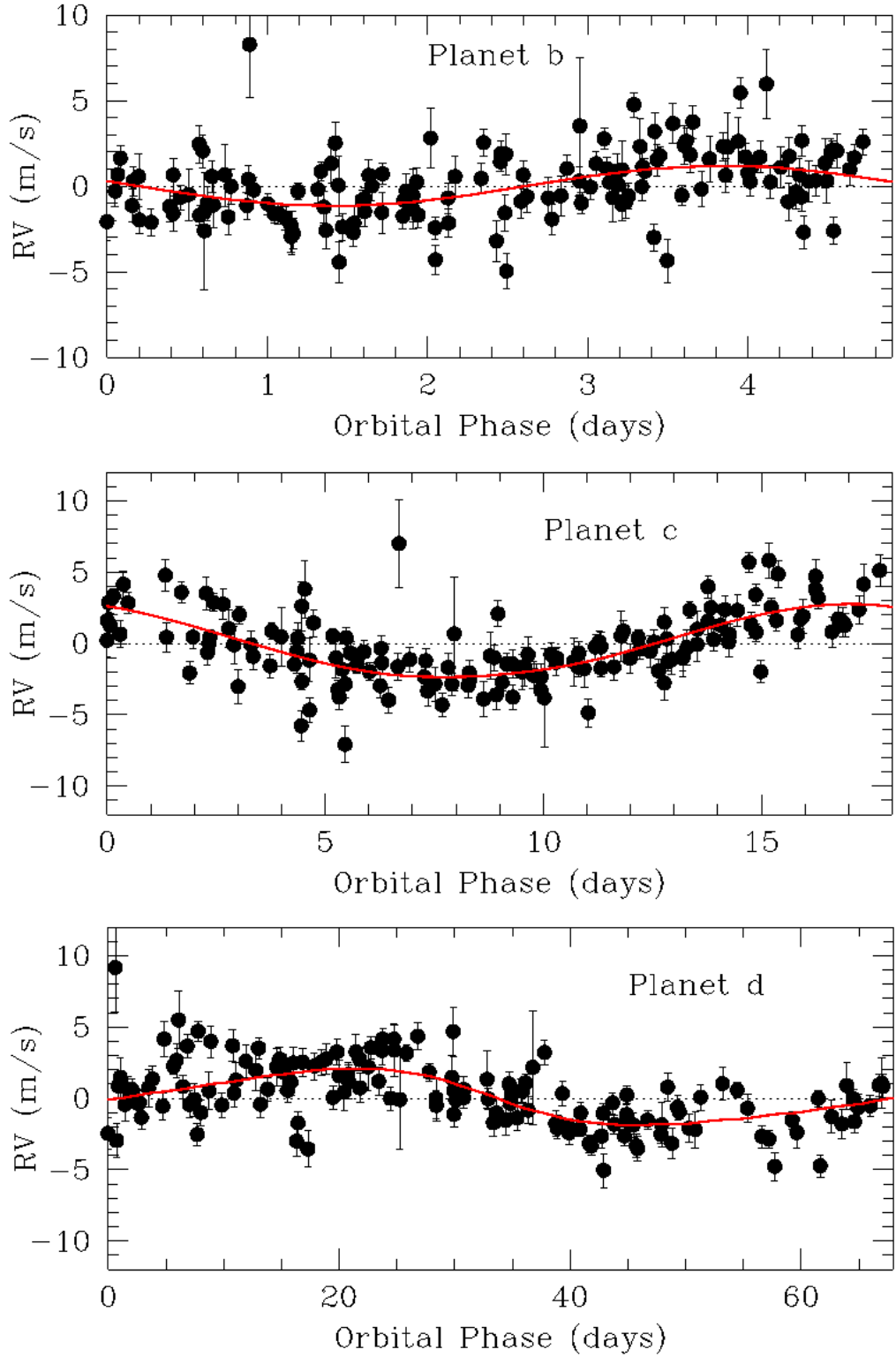}
\caption{Phased velocity signal of each planet with the other planets 
removed.  All three signals are well-sampled in phase.}
\label{phaseplots}
\end{figure}

\clearpage

\begin{deluxetable}{lllllll}
\rotate
\tabletypesize{\scriptsize}
\tablecolumns{7}
\tablewidth{0pt}
\tablecaption{Wolf 1061 Planetary System Parameters }
\tablehead{
\colhead{Parameter} & \multicolumn{3}{c}{Eccentricity Free} &
\multicolumn{3}{c}{Circular Orbits} \\
\colhead{} & \colhead{Wolf 1061b} & \colhead{Wolf 1061c} &
\colhead{Wolf 1061d} & \colhead{Wolf 1061b} & \colhead{Wolf 1061c} &
\colhead{Wolf 1061d} }

\startdata
\label{planetparams}
Period (days) & 4.8876$\pm$0.0014 & 17.867$\pm$0.011 & 67.27$\pm$0.12 & 4.8871$\pm$0.0011 & 17.870$\pm$0.005 & 67.28$\pm$0.08 \\
m sin $i$ (\Mearth) & 1.36$\pm$0.23 & 4.25$\pm$0.37 & 5.21$\pm$0.68 & 1.33$\pm$0.21 & 4.10$\pm$0.31 & 4.97$\pm$0.58 \\
$K$ (\ms) & 1.29$\pm$0.22 & 2.70$\pm$0.25 & 2.24$\pm$0.30 & 1.27$\pm$0.20 & 2.53$\pm$0.19 & 1.97$\pm$0.23 \\
Eccentricity & 0.0 (fixed) & 0.19$\pm$0.13 & 0.32$\pm$0.16 & 0.0 (fixed) & 0.0 (fixed) & 0.0 (fixed) \\
$\omega$ (degrees) & \nodata & 37$\pm$47 & 102$\pm$70 & \nodata & \nodata & \nodata \\
Mean anomaly (degrees) & 61$\pm$41 & 313$\pm$59 & 170$\pm$41 & 279$\pm$30 & 191$\pm$20 & 255$\pm$21 \\
$a$ (AU) & 0.035509$\pm$0.000007 & 0.08427$\pm$0.00004 & 0.2039$\pm$0.0002 & 0.035507$\pm$0.000006 & 0.08428$\pm$0.00002 & 0.2040$\pm$0.0002 \\
\hline
$\chi^2_{\nu}$ & 3.72 & & & 3.77 & & \\
RMS of fit & 1.86\,\ms\ & & & 1.87\,\ms\ & & \\
FAP of signal & 0.0052 & $<$0.0001 & $<$0.0001 & & & \\
Transit probability & 14.0\% & 5.9\% & 2.6\% & & & \\
Transit depth (mmag) & 2.6 & 3.3 & 5.2 & & & \\
\enddata
\end{deluxetable}

\clearpage

\begin{deluxetable}{lrlrcrlrlrlrlrl}
\rotate
\tabletypesize{\scriptsize}
\tablecolumns{15}
\tablewidth{0pt}
\tablecaption{Wolf 1061 velocities from this work and HARPS, cross-correlation FWHM, Bisector spans, and the S-index, Na D and H$\alpha$~activity indices. 
This table is published in its entirety in the electronic edition of this paper, a portion is shown here for guidance regarding its form and content.}
\tablehead{
\colhead{Julian Date} & \multicolumn{2}{c}{Velocity} & \multicolumn{2}{c}{HARPS Vel.} & \multicolumn{2}{c}{FWHM} & \multicolumn{2}{c}{BIS. } & \multicolumn{2}{c}{S-index} & \multicolumn{2}{c}{Na\,D index} & \multicolumn{2}{c}{H$\alpha$ index} \\

\colhead{} &       \colhead{RV} & \colhead{error}       &        \colhead{RV+21035} & \colhead{error}        &         \colhead{FWHM} & \colhead{error}       &         \colhead{BIS} & \colhead{error}        &            \colhead{$S$} & \colhead{error}             & \colhead{NaD} & \colhead{error} &        \colhead{H$\alpha$} & \colhead{error} \\

\colhead{} &       \colhead{\ms} & \colhead{\ms}       &        \colhead{\ms} & \colhead{\ms}        &         \colhead{\ms} & \colhead{\ms}       &         \colhead{\ms} & \colhead{\ms}        &            \colhead{} & \colhead{}             & \colhead{} & \colhead{} &        \colhead{} & \colhead{} \\

}
\startdata
\label{starRVs}
2453158.68022 &  0.87 & 1.15 &  2.69 & 0.89 & 3415 & 22 &  -3.2 & 2.2 &  0.0588 & 0.0045 &  5.697 & 0.037 & 0.077771 & 0.000084 \\ 
2453203.58361 & -4.48 & 0.75 & -1.50 & 0.52 & 3405 & 22 &  -1.2 & 1.5 &  0.0635 & 0.0011 &  4.924 & 0.028 & 0.076379 & 0.000075 \\ 
2453484.84222 & -2.74 & 0.61 & -0.51 & 0.41 & 3394 & 22 &   0.0 & 1.2 &  0.0625 & 0.0010 &  5.015 & 0.024 & 0.076899 & 0.000063 \\ 
2454172.85054 &  2.92 & 0.57 &  1.02 & 0.39 & 3398 & 23 &  -4.0 & 1.1 &  0.0806 & 0.0010 &  5.622 & 0.024 & 0.077449 & 0.000061 \\ 
2454293.65140 & -4.72 & 0.63 & -0.22 & 0.42 & 3399 & 21 &  -1.7 & 1.2 &  0.0684 & 0.0010 &  4.896 & 0.024 & 0.076643 & 0.000065 \\ 
2454340.56988 &  0.12 & 0.49 &  3.05 & 0.34 & 3397 & 22 &  -0.8 & 1.0 &  0.0736 & 0.0009 &  5.027 & 0.020 & 0.077062 & 0.000053 \\ 
2454342.50174 & -4.50 & 0.48 & -1.63 & 0.33 & 3400 & 23 &  -0.4 & 0.9 &  0.0766 & 0.0009 &  5.129 & 0.020 & 0.077502 & 0.000053 \\ 
2454343.50450 & -5.79 & 0.65 & -3.13 & 0.44 & 3401 & 22 &  -2.5 & 1.3 &  0.0753 & 0.0011 &  5.184 & 0.025 & 0.077544 & 0.000068 \\ 
2454344.51023 & -3.20 & 0.76 & -2.33 & 0.52 & 3405 & 23 &   2.6 & 1.5 &  0.0772 & 0.0012 &  5.134 & 0.028 & 0.077435 & 0.000076 \\ 
2454345.48199 & -3.26 & 0.67 & -0.25 & 0.45 & 3406 & 22 &  -1.0 & 1.3 &  0.0832 & 0.0012 &  5.441 & 0.026 & 0.078085 & 0.000069 \\ 
\enddata
\end{deluxetable}

\end{document}